\documentclass[english]{article}
\usepackage[T1]{fontenc}
\usepackage[latin9]{inputenc}
\usepackage[active]{srcltx}
\usepackage{amsmath}
\usepackage{amssymb}
\usepackage{setspace}
\PassOptionsToPackage{normalem}{ulem}
\usepackage{ulem}
\makeatletter
\newcommand{\lyxaddress}[1]{
\par {\raggedright #1
\vspace{1.4em}
\noindent\par}
}

\makeatother

\usepackage{babel}
\begin{document}

\title{Quaternion Octonion Reformulation of Grand Unified Theories}

\author{Pushpa$^{(1)}$, P. S. Bisht$^{(1,2)}$, Tianjun Li$^{(2)}$ and
O. P. S. Negi$^{(1.2)}$ }

\maketitle

\lyxaddress{\begin{center}
1.\textsuperscript{}Department of Physics, Kumaun University, S.
S. J. Campus, Almora - 263601 (U.K.), India
\par\end{center}}

\lyxaddress{\begin{center}
2. Institute of Theoretical Physics,Chinese Academy of Sciences,
Zhong Guan Cun East Street 55, P. O. Box 2735, Beijing 100190, P.
R. China.
\par\end{center}}

\lyxaddress{\begin{center}
Email-pushpakalauni60@yahoo.co.in; ps\_bisht 123@rediffmail.com;
tli@itp.ac.cn; ops\_negi@yahoo.co.in
\par\end{center}}
\begin{abstract}
In this paper, Grand Unified theories are discussed in terms of quaternions
and octonions by using the relation between quaternion basis elements
with Pauli matrices and Octonions with Gell Mann $\lambda$ matrices.
Connection between the unitary groups of GUTs and the normed division
algebra has been established to re-describe the $SU(5)$ gauge group.
We have thus described the $SU(5)$ gauge group and its subgroup $SU(3)_{C}\times SU(2)_{L}\times U(1)$
by using quaternion and octonion basis elements. As such the connection
between $U(1)$ gauge group and complex number, $SU(2)$ gauge group
and quaternions and $SU(3)$ and octonions is established. It is concluded
that the division algebra approach to the the theory of unification
of fundamental interactions as the case of GUTs leads to the consequences
towards the new understanding of these theories which incorporate
the existence of magnetic monopole and dyon.


\end{abstract}

\section{Introduction}

In spite of all the successes of the Standard Model (SM) of elementary
particles, it leaves many unresolved questions to be considered as
the complete theory of matter. The next attempt is described as a
grand unified theory (GUT) with more symmetry and reduces to the Standard
Model at lower energies. It is thus considered as an attempt to describe
the physics at higher energies where the three gauge interactions
(namely the electromagnetic, weak, and strong interactions) of the
Standard Model are merged into one single interaction characterized
by one larger gauge symmetry with one unified coupling constant \cite{key-1}.
Here the standard model gauge groups $SU(3)\times SU(2)\times U(1)$
are combined into a single simple gauge group i.e. $SU(5)$. A series
of hypothesis were argued \cite{key-2,key-3,key-4,key-5} in order
to leading to the conclusion that a gauge theory based on the $SU(5)$
gauge group provide a unified description of the strong and electroweak
interaction. The Georgi \textendash{} Glashow model \cite{key-1}
was preceded by the Semi simple Lie algebra of Pati \textendash{}
Salam model \cite{key-6} to unify gauge interactions. On the other
hand, the four division algebras ($\mathbb{R}$ (real numbers), $\mathbb{C}$
(complex numbers), $\mathbb{H}$ (quaternions) and $\mathcal{O}$
(octonions)) also play an important role in physics and mathematics
particularly for unification program of fundamental interactions.
It is remarkable that the existence of all three exception algebras
and sub algebras is mathematically motivated by their connection through
the maximum division algebra i.e. algebra of octonions \cite{key-7}.
Octonions are widely used for the understanding of unification structure
of  successful gauge theory of fundamental interaction. Recently,
we \cite{key-8} have made an attempt to develop the quaternionic
formulation of Yang\textendash{}Mill\textquoteright{}s field equations
and octonion reformulation of quantum chromo dynamics (QCD) by taking
magnetic monopoles \cite{key-9,key-10,key11} and dyons (particles
carrying electric and magnetic charges) \cite{key-12,key-13,key-14}
into account. It has been shown that the three quaternion units explain
the structure of Yang-Mill\textquoteright{}s field while the seven
octonion units provide the consistent structure of $SU(3)_{C}$ gauge
symmetry of quantum chromo dynamics. Here we apply entirely different
approach for the quaternion gauge theory of electroweak interactions
and octonion gauge structure for quantum chromo dynamics (QCD) compare
to the approaches adopted earlier by Morita \cite{key-15,key-16}
and others \cite{key-17,key-18}. It is already explored \cite{key-8}
that the three quaternion units explain the structure of Yang- Mill\textquoteright{}s
field while the seven octonion units provide the consistent structure
of $SU(3)_{C}$ gauge symmetry of quantum chromo dynamics (QCD) as
these are connected with the well known $SU(3)$ Gellmann $\lambda$
matrices. Keeping these facts in mind and extending our previous work
\cite{key-8}, in this paper, we have made an attempt to discuss the
Grand Unified Theories and then to reformulate them in terms of algebra
of quaternions and octonions in simple and consistent manner. Accordingly,
the connection between the unitary groups of GUTs and the normed division
algebra has been established to re-describe the $SU(5)$ gauge group
and its constituents. So, we have reformulated the GUTs gauge group
$SU(5)$ its subgroup $SU(3)_{C}\times SU(2)_{L}\times U(1)$ in terms
of quaternion and octonion basis elements. Hence, the connection between
$U(1)$ gauge group and complex number, $SU(2)$ gauge group and quaternions
and $SU(3)$ and octonions has been re-established. It is shown that
grand unified theories in terms of quaternion and octonion contain
the magnetic monopole. It is concluded that the division algebra approach
to the the theory of unification of fundamental interactions as the
case of GUTs leads to the consequences towards the new understanding
of these theories which incorporate the existence of magnetic monopole
and dyon. Three different imaginaries associated octonion formulation
may be identified with three different colors (red, blue and green)
while the Gell Mann Nishijima $\lambda$ are described in terms of
simple and compact notations of octonion basis elements. The symmetry
breaking mechanism of non - Abelian gauge theories in terms of quaternion
and octonion opens the window towards the discovery of two type of
gauge bosons associated with electric and magnetic charges.

\section{SU(5) Gauge Symmetry}

\begin{spacing}{1.5}
Grand Unified Theories \cite{key-19} are based on the mathematical
symmetry group $SU(5)$. The gauge group of the Glashow - Salam -
Weinberg theory $SU(2)\times U(1)$ and the $SU(3)$ group of the
strong interaction are the constituents of a larger symmetry. The
simplest group that incorporates the product $SU(3)$$_{C}$$\times$$SU(2)_{L}$$\times$$U(1)$
as a subgroup is $SU(5)$. An arbitrary unitary matrix can be represented
in terms of an exponential of a Hermitian matrix $H$ as \cite{key-20},
\begin{align}
\widehat{U}=e^{i\hat{H}}\:,\: & \widehat{H}^{\dagger}=H;\label{eq:1}
\end{align}
where $\hat{H}$ is known as the generating matrix for $\hat{U}$
, which can be written as,

\begin{align}
\hat{U}=\exp\left(i\delta\widehat{H}\right)\thickapprox & 1+i\delta\widehat{H}.\label{eq:2}
\end{align}
The multiplication of two matrices $U_{1}$, $U_{2}$ corresponds
to the sum of the infinitesimal Hermitian matrices as,
\end{spacing}

\begin{align}
\widehat{U_{2}}\widehat{U_{1}}\approx & 1+i\left(\delta\hat{H_{2}}+\delta\hat{H_{1}}\right)\label{eq:3}
\end{align}
where quadratic terms are neglected. A complete set of linearly independent
Hermitian matrices is termed as a set of generators for the unitary
matrices. The unitary condition $UU^{\dagger}=1$ and the uni modular
condition $\det U=1$ leave the $5^{2}-1=24$ independent matrices
for $SU(5)$ symmetry. $U$ may then be described as,

\begin{spacing}{1.5}
\begin{align}
U= & \exp\left(-i\sum_{a=1}^{24}A_{\mu}^{a}L^{a}\right);\label{eq:4}
\end{align}
where $L^{a}$ contains $24$ generators which are Hermitian and traceless
and $A_{\mu}^{a}$ is a $5\times5$ matrix. Thus, the $SU(5)$ symmetry
splits into the $SU(3)$ symmetry of strong force along with the $SU(2)\times U(1)$
gauge symmetry of electroweak force. The $SU(2)\times U(1)$ gauge
symmetry also splits into a $SU(2)$ sub symmetry of the weak interaction
and the $U(1)$ sub symmetry of the electromagnetic interaction. Here
it is to be cleared that the $5\times5$ matrices $L$ is taken in
such a way that the colour group $SU(3)$ acts on first three rows
and columns in terms of octonions \cite{key-8}, while the $SU(2)$
group operates on the last two rows and columns by means of quaternions
\cite{key-8} and $U(1)$ is the singlet associated with complex numbers.
Thus the algebra of $SU(5)$ illustrates as $\mathcal{O}$$\oplus$
$\mathbb{\mathbb{Q}}$ $\oplus$ $\mathbb{C}$. $\mathcal{O}$ is
used for octonions, $\mathbb{Q}$ for quaternions and $\mathbb{C}$
for complex numbers. $\mathcal{O}$ has the connection with $SU(3)$,
$\mathbb{Q}$ describes $SU(2)$ and $\mathbb{C}$ is linked with
$U(1)$. This gives the $SU(3)\times SU(2)\times U(1)$ subgroup structure
of $SU(5)$ in terms of the constituents of three division algebras
namely octonions $\mathcal{O}$, quaternion $\mathbb{Q}$ and the
algebra of complex numbers $\mathbb{C}$.
\end{spacing}

\section{Quaternion-Octonion Reformulation of $SU(5)$ gauge Symmetry}

\begin{spacing}{1.5}
Let us break up a $5\times5$ square matrix $SU(5)$  into four blocks,
consisting of two smaller squares and two rectangles. The upper left
- hand corner block denotes a $3\times3$ matrix while the lower right-hand
corner block describes a $2\times2$ matrix. Off-diagonal upper right
- hand and lower left - hand corners consist respectively the $3\times2$
and $2\times3$ rectangular matrices. The generators are described
in such a way that the first eight generators are associated with
the generators of $SU(3)$ symmetry as

\begin{align}
L_{a}= & \left(\begin{array}{cc}
\lambda^{a} & 0\\
0 & 0
\end{array}\right);\label{eq:5}
\end{align}
where $a=1,\,2,\,....\,8$  and $\lambda$ are the well known $3\times3$
Gell Mann matrices. Here replace \cite{key-8} Gell Mann $\lambda$
matrices by octonion basis elements. For the $9^{th}$, $10^{th}$
and $11^{th}$ generators, we may use quaternion basis elements in
the $2\times2$ block which is related to Pauli matrices $\sigma_{j}$.
Let us write the quaternion scalar part as $1$ corresponding to the
$2\times2$ unit matrix and three imaginary quaternion units $e_{1},\, e_{2},\, e_{3}$
are connected with Pauli - spin matrices as
\begin{eqnarray}
e_{0}= & 1;\,\,\,\,\, & e_{j}=-i\sigma_{j}.\label{eq:6}
\end{eqnarray}
So, we may write

\begin{align}
L^{8+j}= & \left(\begin{array}{cc}
0 & 0\\
0 & ie_{j}
\end{array}\right);\label{eq:7}
\end{align}
where $j=1,2,3$ belonging to the three generators of $SU(2)$ gauge
group. Accordingly the $L^{12}$ describes the hyper charge corresponding
to $U(1)$ gauge group associated with the scalar part of a quaternion
(or complex), so that

\begin{align}
L^{12}= & \frac{1}{\sqrt{15}}diag\left(-2,-2,-2,3,3\right)=\frac{1}{\sqrt{15}}\left[\begin{array}{cc}
-2I_{3} & 0\\
0 & 3I_{3}
\end{array}\right].\label{eq:8}
\end{align}
For the next set of matrices, it will be convenient to define rectangular
matrices $A$ and $B$ \cite{key-21} as,
\end{spacing}

\begin{align}
A_{1}=\left(\begin{array}{cc}
1 & 0\\
0 & 0\\
0 & 0
\end{array}\right);\,\,\,\, & A_{2}=\left(\begin{array}{cc}
0 & 0\\
1 & 0\\
0 & 0
\end{array}\right);\,\,\,\,\, A_{3}=\left(\begin{array}{cc}
0 & 0\\
0 & 0\\
1 & 0
\end{array}\right);\label{eq:9}
\end{align}
and

\begin{align}
B_{1}=\left(\begin{array}{cc}
0 & 1\\
0 & 0\\
0 & 0
\end{array}\right);\,\,\,\, & B_{2}=\left(\begin{array}{cc}
0 & 0\\
0 & 1\\
0 & 0
\end{array}\right);\,\,\,\,\, B_{3}=\left(\begin{array}{cc}
0 & 0\\
0 & 0\\
0 & 1
\end{array}\right).\label{eq:10}
\end{align}
Thus, the $13^{th}$ to $24^{th}$ generators of $SU(5)$ symmetry
are expressed as,

\begin{align}
L^{13,15,17}= & L^{11+2k}=e_{0}\left(\begin{array}{cc}
0 & A_{k}\\
A_{k}^{T} & 0
\end{array}\right)=e_{0}L^{11+2k};\nonumber \\
L^{14,16,18}= & L^{12+2k}=-e_{3}L^{11+2k};\nonumber \\
L^{19,21,23}= & L^{17+2k}=e_{0}\left(\begin{array}{cc}
0 & B_{k}\\
B_{k}^{T} & 0
\end{array}\right)=e_{0}L^{17+2k};\nonumber \\
L^{20,22,24}= & L^{18+2k}=e_{3}L^{17+2k};\label{eq:11}
\end{align}
where $k=1,2,3.$ Hence we may define the precise association between
the vector gauge bosons of $SU(5)$ GUTs , vector gauge bosons of
the electro-weak model $(W_{\mu}^{\pm},\, W_{\mu}^{3},\, B_{\mu})$
and the eight vector gluon G$_{\mu}^{a}$ of QCD as,

\begin{align}
A_{\mu}^{1,2,.....8}\Longrightarrow & G_{\mu}^{a}\,(\forall a=1,2,...,8);\nonumber \\
A_{\mu}^{9,10}\Longrightarrow & W^{\pm};\nonumber \\
A_{\mu}^{11}\Longrightarrow & W_{\mu}^{3};\nonumber \\
A_{\mu}^{12}\Longrightarrow & B_{\mu}.\label{eq:12}
\end{align}
The new vector bosons, which are specified only by $SU(5)$ and do
not contribute to the Standard Model, are then be expressed as,

\begin{align}
A_{\mu}^{13,14.......18}= & X_{\mu}^{b}\,(\forall b=1,2,...,6);\nonumber \\
A_{\mu}^{19,20,......24}= & Y_{\mu}^{b}\,\,(\forall b=1,2,...,,6).\label{eq:13}
\end{align}
Generalizing equations (\ref{eq:11} to \ref{eq:13}), we may write
the $5\times5$ matrix representation for the $SU(5)$ gauge field
$A_{\mu}$in a compact form as,

\begin{spacing}{1.5}
\begin{align}
A_{\mu}= & \sqrt{2}\sum_{i=1}^{24}A_{\mu}^{i}L^{i};\label{eq:14}
\end{align}
Which is also expressed as,

\begin{align}
A_{\mu}^{i}= & \left(\begin{array}{cc}
P & Q\\
R & S
\end{array}\right);\label{eq:15}
\end{align}
where
\begin{align}
P=G_{\mu}-\frac{2}{\sqrt{3}}B_{\mu}I\,\,; & \,\, Q=\left(\begin{array}{cc}
\overline{X}^{1} & \overline{Y}^{1}\\
\overline{X}^{2} & \overline{Y}^{2}\\
\overline{X}^{3} & \overline{Y}^{3}
\end{array}\right);\label{eq:16}
\end{align}
and

\begin{align}
R=\left(\begin{array}{ccc}
X_{1} & X_{2} & X_{3}\\
Y_{1} & Y_{2} & Y_{3}
\end{array}\right)\,\,; & \,\, S=\left(\begin{array}{cc}
\frac{W_{3}}{\sqrt{2}}+\frac{3B}{\sqrt{30}} & W^{+}\\
W^{-} & \frac{-W_{3}}{\sqrt{2}}+\frac{3B}{\sqrt{30}}
\end{array}\right).\label{eq:17}
\end{align}
The symmetry breaking of $SU(5)$ into $SU(3)\times SU(2)\times U(1)$
can be done in a similar way as the breaking of $SU(3)$ into $SU(2)\times U(1)$.
The $S$ generator in the adjoint representation of $SU(5)$ commutes
with $SU(3)\times SU(2)\times U(1)$. The $24$ adjoint representations
with vacuum values in the $S$ direction will consequently break $SU(5)$
into $SU(3)\times SU(2)\times U(1)$. Under $SU(3)\times SU(2)\times U(1)$
the decomposition of the adjoint $24$ representation may then be
expressed as,
\end{spacing}

\begin{align}
[24]_{5}= & (8,\,1)+(\overline{3},\,2)+(3,\,2)+(1,\,3)+(1,\,1).\label{eq:18}
\end{align}
Here, the gauge bosons for strong interactions due to the exchange
of quarks and colors are the eight gluon associated with the color
gauge group $SU(3)_{C}$ and for the electro weak interactions $SU(2)\times U(1)$
are the intermediate Weak bosons and photons. Following arguments
be made accordingly as
\begin{itemize}
\begin{spacing}{1.5}
\item There is a $SU(3)_{C}$ octet of gluon $G_{\mu}^{a}=\left(1,\,2,.....\,8\right)$$\longmapsto\left(8,\,1\right)$.\end{spacing}

\item There is an isovector of intermediate bosons , $W_{\mu}^{i}\,(i=1,\,2,\,3)\longmapsto\left(1,\,3\right)$.
\item There is an isoscalar field of the hyper charged boson $B_{\mu}\longmapsto\left(1,\,1\right)$.
\end{itemize}
\begin{spacing}{1.5}
In addition, to the above three arguments there are twelve more gauge
bosons, belonging to the representations $\left(3,\,2\right)$ and
$\left(\overline{3},\,2\right)$. These gauge bosons form an isospin
doublet of bosons and their antiparticles, which are colored. The
simplest representation of $SU(5)$ is the five dimensional fundamental
one $\psi_{5}$ which may be represented by a column matrix as

\begin{align}
\psi_{5}= & \left(\begin{array}{c}
a^{1}\\
a^{2}\\
a^{3}\\
a^{4}\\
a^{5}
\end{array}\right).\label{eq:19}
\end{align}
In the $SU(2)$ symmetry of weak interactions the quaternions occur
only on the rows $4$ and $5$ and we see that $a_{1}$,$\, a_{2}$
and $a_{3}$ are unaffected by the operation of $SU(2)$ generators
i.e. quaternions. For the case of $SU(5)$, the covariant derivative
may be written in terms of matrix representation as,

\begin{align}
D_{\mu}= & \partial_{\mu}-\frac{ig}{2}A_{\mu}.\label{eq:20}
\end{align}
Then all $24$ gauge bosons $A_{\mu}^{i}$ $(i=1,...,24)$ are conveniently
represented by a $5\times5$ matrix. Adopting  these $24$ generators
of $SU(5)$, we may write the equation $\left(\ref{eq:14}\right)$
in the following form,

\begin{align}
A_{\mu}=\sqrt{2} & \sum_{a=1}^{24}A_{\mu}^{a}L_{a}=\left[\sum_{a=1}^{8}G_{\mu}^{a}L_{a}+\sum_{a=9}^{11}A_{\mu}^{a}L_{a}+B_{\mu}L_{12}+\sum_{a=13}^{24}A_{\mu}^{a}L_{a}\right].\label{eq:21}
\end{align}
where $\sum_{a=1}^{8}\, G_{\mu}^{a}L_{a}$ has been expressed in terms
of octonion units as,

\begin{align}
\sum_{a=1}^{8}G_{\mu}^{a}L_{a}= & \sum_{a=1}^{7}G_{\mu}^{a}e_{a}+G_{\mu}^{8}e_{3};\label{eq:22}
\end{align}
while the second term containing $L_{9},\, L_{10},\, L_{11}$ are
defined in terms of quaternion units as
\begin{align}
\sum_{a=1}^{8}\lambda_{a}\alpha^{a}\left(x\right)= & -i\sum_{q=1}^{7}e_{q}\beta^{q}\left(x\right);\label{eq:23}
\end{align}
whereas the term $-i\sum_{q=1}^{7}e_{q}\beta^{q}\left(x\right)$ is
expressed in terms of the matrix as,

\begin{align}
-i\sum_{q=1}^{7}e_{q}\beta^{q}\left(x\right)= & \left[\begin{array}{ccc}
\alpha_{3}+\frac{\alpha_{8}}{\sqrt{3}} & \alpha_{1}-i\alpha_{2} & \alpha_{4}-i\alpha_{5}\\
\alpha_{1}+i\alpha_{2} & -\alpha_{3}+\frac{\alpha_{8}}{\sqrt{3}} & \alpha_{6}-i\alpha_{7}\\
\alpha_{4}+i\alpha_{5} & \alpha_{6}+i\alpha_{7} & -\frac{2\alpha_{8}}{\sqrt{3}}
\end{array}\right].\label{eq:24}
\end{align}
Hence we may write equation (\ref{eq:21}) as

\begin{align}
A_{\mu}= & \left(\begin{array}{cc}
-i\sum_{q=1}^{7}e_{q}\beta^{q}\left(x\right) & B\\
C & -i\sum_{i=1}^{3}W_{\mu}^{i}e_{i}
\end{array}\right)+\frac{B_{\mu}}{2\sqrt{15}}diag\left(-2,-2,-2,\,3,\,3\right);\label{eq:25}
\end{align}
which is further simplified to the following form in terms of $5\times5$
matrix on substituting the value of $P,\, Q,\, R,\, S$ i.e.

\begin{align}
A_{\mu}= & \left(\begin{array}{ccccc}
\alpha_{3}+\frac{\alpha_{8}}{\sqrt{3}}-\frac{2B_{\mu}}{\sqrt{15}} & \alpha_{1}-i\alpha_{2} & \alpha_{4}-i\alpha_{5} & \overline{X}^{1} & \overline{Y}^{1}\\
\alpha_{1}+i\alpha_{2} & -\alpha_{3}+\frac{\alpha_{8}}{\sqrt{3}}-\frac{2B_{\mu}}{\sqrt{15}} & \alpha_{6}-i\alpha_{7} & \overline{X}^{2} & \overline{Y}^{2}\\
\alpha_{4}+i\alpha_{5} & \alpha_{6}+i\alpha_{7} & -\frac{2\alpha_{8}}{\sqrt{3}}-\frac{2B_{\mu}}{\sqrt{15}} & \overline{X}^{3} & \overline{Y}^{3}\\
X_{1} & X_{2} & X_{3} & W_{\mu}^{3}+\frac{3B_{\mu}}{\sqrt{15}} & W_{\mu}^{1}-iW_{\mu}^{2}\\
Y_{1} & Y_{2} & Y_{3} & W_{1}^{\mu}+iW_{\mu}^{2} & -W_{\mu}^{3}+\frac{3B_{\mu}}{\sqrt{15}}
\end{array}\right).\label{eq:26}
\end{align}
As such, the covariant derivative (\ref{eq:20}) associated with $W_{\mu}^{3}$
and $B_{\mu}$ may then be expressed in terms of the coupling of $A_{\mu}$
and $Z_{\mu}$ by extracting the $11^{th}$ and $12^{th}$ generators
of covariant derivative \cite{key-22,key-23} as,

\begin{align}
D_{\mu}= & \partial_{\mu}-i\frac{g}{2}\left(W_{\mu}^{3}L^{11}+B_{\mu}L^{12}\right)\nonumber \\
= & \partial_{\mu}-i\frac{g}{2}\left[A_{\mu}\left(\sin\theta_{W}L^{11}+\cos\theta_{W}L^{12}\right)+Z_{\mu}\left(\cos\theta_{W}L^{11}-\sin\theta_{W}L^{12}\right)\right]\label{eq:27}
\end{align}
 where $\theta_{W}$ is defined as the Weinberg angle.
\end{spacing}

\section{Discussion and Conclusion }

\begin{spacing}{1.5}
We have already shown earlier \cite{key-8} that the quaternion and
octonion gauge theories contain the magnetic monopole. As such, the
reformulation of grand unified theories (GUTs) in terms of the gauge
group $SU(5)$ and its splitting to $SU(3)_{C}\times SU(2)\times U(1)$
may lead to the simultaneous existence of two different gauge theories
associated with electric and magnetic charges ( i.e. dyons). This
this approach may be used as the milestone for the unification of
fundamental interaction at one end and the existence of magnetic monopole
at other end so that the unanswered question for the existence of
magnetic monopoles can be tackled. Three different imaginaries responsible
for the creation of octonion formulation may be identified as the
three different colors (red, blue and green) while the matrix form
of Gell Mann Nishimijja matrices establish well defined connection
with seven octonion elements. The symmetry breaking mechanism of non
- Abelian gauge theory in terms of quaternion octonion leads to the
existence of massive gauge bosons associated with electric and magnetic
charges. The present theory may also provide a variety of compiled
phenomenon involving the various particles and forces such as
\end{spacing}
\begin{itemize}
\begin{spacing}{1.5}
\item The occurrence of particle transition between number of family such
as not only between electron and electron neutrino or u, d quarks
but also with a monopole and monopole neutrino with monopole quarks.
One can also relate the quarks and neutrinos with the dyons.
\item There exists two type of gauge bosons (W and Z) due to electron and
monopole in order to transit forces.
\item The fact that quarks never appear either alone (a phenomenon called
quark confinement) or in combination which have a net color charge.
Here we may emphasize that if quarks are considered as dyons the problem
may be resolved automatically.
\item There exists two types of gluon which transit the strong forces are
capable of changing the ``color'' or chromo magnetic color charges
of quarks as dyons which is in agreement with the results of Kühne
\cite{key-24}. \end{spacing}

\end{itemize}
As such, the foregoing analysis describes the embedding of $U(1)$$\times$$U(1)$
model in grand unified theories (GUTs) so that one may imagine the
underlying group to be $SU(5)$$\times$ $SU(5)$ where second $SU(5)$
describes the hypothetical magnetic photons, chromo- magnetic gluons
and (iso-) magnetic W, Z, X and Y bosons.

\textbf{ACKNOWLEDGMENT}: Two of us (PSB \&OPSN) are thankful to UNESCO
and Third World Academy of Sciences, Trieste (Italy) for providing
them UNESCO-TWAS Associateship. The hospitality and research facilities\textbf{
}provided by ProfessorYue-Liang Wu\textbf{, }Director \textbf{I}TP,
in Institute of Theoretical Physics and Kavli Institute of Theoretical
Physics at Chinese Academy of Sciences,\textbf{ }Beijing (China) under
the frame work of TWAS Associate Membership Scheme are also acknowledged.


\begin{thebibliography}{10}
\bibitem[1]{key-1} H. Georgi and S. Glashow, \textbf{``Unity of
All Elementary-Particle Forces''}, Phys. Rev. Lett., \textbf{\uline{32}}
(1974) 438.

\bibitem[2]{key-2} Graham G. Ross, \textbf{``Grand Unified Theories}'',
Benjamin/Cummings, 1985.

\bibitem[3]{key-3} R. N. Mohapatra, \textbf{``Unification and Supersymmetry:
The Frontiers of Quark-Lepton Physics}'', Springer, 1992.

\bibitem[4]{key-4} E. Witten, \textbf{``Grand unification with and
without supersymmetry, in Introduction to supersymmetry in particle
and nuclear physics}'', Editors. O. Castanos, A. Frank, L. Urrutia,
Plenum Press, 1984, pp. 53\textendash{}76.

\bibitem[5]{key-5} J. Baez and J. Huerta, ``\textbf{The Algebra
of Grand Unified Theories'', }Bull. Am. Math.Soc., \textbf{47} (2010)
483: eprint-arXiv:0904.1556 {[}hep-th{]}.

\bibitem[6]{key-6}  J. Pati and A. Salam, \textbf{``Lepton Number
as the Fourth Color''}, Physical Review, \textbf{\uline{D10}}
(1974), 275.

\bibitem[7]{key-7} J. C. Baez, \textbf{``The Octonions''}, Bull.
Amer. Math. Soc., \textbf{\uline{39}} (2001), 145.

\bibitem[8]{key-8} Pushpa, P. S. Bisht, Tianjun Li and O. P. S. Negi,
\textbf{``Quaternion Octonion Reformulation of Quantum Chromodynamics}'',
Int. J. Theor. Phys., 50 (2011), 594.

\bibitem[9]{key-9} P. A. M. Dirac, \textbf{``Quantised Singularities
in the Electromagnetic Field''}, Proc. Royal Society, \textbf{\uline{A133}}
(1931), 60.

\bibitem[10]{key-10} G. 't Hooft, \textbf{``Magnetic monopoles in
unified gauge theories''}, Nucl. Phys., \textbf{\uline{B79}} (1974),
276.

\bibitem[11]{key11} A. M. Polyakov,\textbf{ ``Particle spectrum
in quantum field theory''}, JETP Lett., \textbf{\uline{20}} (1974),
194.

\bibitem[12]{key-12} J. Schwinger, ``\textbf{Dyons Versus Quarks''},
Science, \textbf{\uline{166}} (1969), 690.

\bibitem[13]{key-13} D. Zwanzinger,\textbf{ ``Quantum Field Theory
of Particles with Both Electric and Magnetic Charges''}, Phys. Rev.,\textbf{\uline{
176 }}(1968), 1489.

\bibitem[14]{key-14} B. Julia and A. Zee , \textbf{``Poles with
both magnetic and electric charges in non-Abelian gauge theory''},
Phys. Rev., \textbf{\uline{D11}} (1975), 2227.

\bibitem[15]{key-15} K. Morita, \textbf{``Octonions, Quarks and
QCD''}, Prog. Theor. Phys., \textbf{\uline{65}} (1981),787.

\bibitem[16]{key-16} K. Morita, \textbf{``An Algebraic Explanation
for the Family Structure of Quarks and Leptons''}, Prog. Theor. Phys.,
\textbf{\uline{66}} (1981), 1519.

\bibitem[17]{key-17} M. Günaydin and F. Gürsey,\textbf{ ``Quark
structure and octonions''}, J. Math. Phys., \textbf{\uline{14}}
(1973), 1651.

\bibitem[18]{key-18} G. M. Dixon, \textbf{``Division Algebras::
Octonions Quaternions Complex Numbers and the Algebraic Design of
Physics''}, Springer (1994).

\bibitem[19]{key-19} A. J. Buras, J. R. Ellis, M. K. Gaillard, D.
V. Nanopoulos, \textbf{``Aspects of the grand unification of strong,
weak and electromagnetic interactions''}, Nucl. Phys., \textbf{\uline{B135
}}(1978), 66.

\bibitem[20]{key-20} G. Müller, \textbf{``Gauge theory of weak interactions''},
4$^{th}$ edition, Springer (2009).

\bibitem[21]{key-21} M. Kaku, \textbf{``Quantum field theory''},
Oxford University Press, (1994).

\bibitem[22]{key-22} G. Ross, \textbf{``Grand Unified Theories''},
Westview Press, ISBN 978-0-805-36968-7, (1984).

\bibitem[23]{key-23} H. Georgi, \textbf{``Unified Gauge Theories''},
Proceedings, Theories and Experiments In High Energy Physics, New
York (1975), 329.

\bibitem[24]{key-24} R. W. Kühne, \textbf{``A Model of Magnetic
Monopoles''}, Mod. Phys. Letts., \textbf{\uline{A12 }}(1997),
40.\end{thebibliography}
\end{document}